\documentstyle[amstex,epsfig]{mn}
\newif\ifAMStwofonts
\AMStwofontstrue

\ifoldfss
  \ifCUPmtlplainloaded \else
    \NewTextAlphabet{textbfit} {cmbxti10} {}
    \NewTextAlphabet{textbfss} {cmssbx10} {}
    \NewMathAlphabet{mathbfit} {cmbxti10} {} 
    \NewMathAlphabet{mathbfss} {cmssbx10} {} 
  \fi
  \ifAMStwofonts
    \ifCUPmtlplainloaded \else
      \NewSymbolFont{upmath} {eurm10}
      \NewSymbolFont{AMSa} {msam10}
      \NewMathSymbol{\upi}     {0}{upmath}{19}
      \NewMathSymbol{\umu}     {0}{upmath}{16}
      \NewMathSymbol{\upartial}{0}{upmath}{40}
      \NewMathSymbol{\leqslant}{3}{AMSa}{36}
      \NewMathSymbol{\geqslant}{3}{AMSa}{3E}

    \fi
  \fi
\fi 

\ifnfssone
  \newmathalphabet{\mathit}
  \addtoversion{normal}{\mathit}{cmr}{m}{it}
  \addtoversion{bold}{\mathit}{cmr}{bx}{it}
  \newmathalphabet{\mathbfit} 
  \addtoversion{normal}{\mathbfit}{cmr}{bx}{it}
  \addtoversion{bold}{\mathbfit}{cmr}{bx}{it}
  \newmathalphabet{\mathbfss} 
  \addtoversion{normal}{\mathbfss}{cmss}{bx}{n}
  \addtoversion{bold}{\mathbfss}{cmss}{bx}{n}
  \ifAMStwofonts
    \ifCUPmtlplainloaded \else
      \UseAMStwoboldmath
      \makeatletter
      \new@mathgroup\upmath@group
      \define@mathgroup\mv@normal\upmath@group{eur}{m}{n}
      \define@mathgroup\mv@bold\upmath@group{eur}{b}{n}
      \edef\UPM{\hexnumber\upmath@group}
      \new@mathgroup\amsa@group
      \define@mathgroup\mv@normal\amsa@group{msa}{m}{n}
      \define@mathgroup\mv@bold\amsa@group{msa}{m}{n}
      \edef\AMSa{\hexnumber\amsa@group}
      \makeatother
      \mathchardef\upi="0\UPM19
      \mathchardef\umu="0\UPM16
      \mathchardef\upartial="0\UPM40
      \mathchardef\leqslant="3\AMSa36
      \mathchardef\geqslant="3\AMSa3E
    \fi
  \fi
\fi 

\ifnfsstwo
  \DeclareMathAlphabet{\mathbfit}{OT1}{cmr}{bx}{it}
  \SetMathAlphabet\mathbfit{bold}{OT1}{cmr}{bx}{it}
  \DeclareMathAlphabet{\mathbfss}{OT1}{cmss}{bx}{n}
  \SetMathAlphabet\mathbfss{bold}{OT1}{cmss}{bx}{n}
  \ifAMStwofonts
    \ifCUPmtlplainloaded \else
      \DeclareSymbolFont{UPM}{U}{eur}{m}{n}
      \SetSymbolFont{UPM}{bold}{U}{eur}{b}{n}
      \DeclareSymbolFont{AMSa}{U}{msa}{m}{n}
      \DeclareMathSymbol{\upi}{0}{UPM}{"19}
      \DeclareMathSymbol{\umu}{0}{UPM}{"16}
      \DeclareMathSymbol{\upartial}{0}{UPM}{"40}
      \DeclareMathSymbol{\leqslant}{3}{AMSa}{"36}
      \DeclareMathSymbol{\geqslant}{3}{AMSa}{"3E}
    \fi
  \fi
\fi 

\ifCUPmtlplainloaded \else
  \ifAMStwofonts \else 
    \def\upi{\pi}
    \def\umu{\mu}
    \def\upartial{\partial}
  \fi
\fi

\title[Extended Ly$\alpha$ emission from a DLA at z = 1.93]
{Extended Ly$\alpha$ emission from a damped Ly$\alpha$ absorber \\
at  z = 1.93, and the relation between DLAs and Lyman-break galaxies
} 
\author[J. U. Fynbo, et al.]
       {J. U. Fynbo$^{1,2,3}$, P. M\o ller$^{1,3,4}$ and S. J. Warren$^{5}$ \\
       $^1$European Southern Observatory, Karl-Schwarzschild-Stra\ss e 2, 
D-85748 Garching, Germany\\
       $^2$Institute of Physics and Astronomy, University of \AA rhus, 
DK-8000 \AA rhus C, Denmark\\
       $^3$Space Telescope Science Institute, 3700 San Martin Drive, 
Baltimore, MD21218, USA\\
       $^4$on assignment from the Space Science Department of ESA\\
       $^5$Blackett Laboratory, Imperial College of Science Technology 
and Medicine, Prince
       Consort Rd, London SW7 2BZ, UK}
\date{Accepted 1998 ???.
      Received ???;
      in original form ???}

\pagerange{\pageref{firstpage}--\pageref{lastpage}}
\pubyear{1997}

\begin{document}

\maketitle

\label{firstpage}

\begin{abstract}
The number of damped Ly$\alpha$ absorbers (DLAs) currently known is
about 100, but our knowledge of their sizes and morphologies is still
very sparse as very few have been detected in emission. Here we present
narrow--band and broad--band observations of a DLA in the field of the
quasar pair Q0151+048A (qA) and Q0151+048B (qB). These two quasars have
very similar redshifts z$_{em}$=1.922, 1.937, respectively, and an
angular separation of 3.27 arcsec. The spectrum of qA contains a DLA at
z$_{abs}$ = 1.9342 (close to the emission redshift) which shows an
emission line in the trough, detected at 4$\sigma$. Our narrow--band
image confirms this detection and we find Ly$\alpha$ emission from an
extended area covering 6x3 arcsec$^{2}$, corresponding to 25x12h$^{-2}$
kpc$^{2}$ (q$_0$=0.5, H$_0$ = 100h km s$^{-1}$). The total Ly$\alpha$
luminosity from the DLA is 1.2 x 10$^{43}$ h$^{-2}$ erg s$^{-1}$, which
is a factor of several higher than the Ly$\alpha$ luminosity found
from other DLAs. The narrow--band image also indicates that qB is not
covered by the DLA. This fact, together with the large equivalent width
of the emission line from the Ly$\alpha$ cloud, the large luminosity,
and the $\sim300$ km s$^{-1}$ blueshift relative to the DLA, can plausibly be
explained if qB is the source of ionising photons, lying to the near
side of the DLA at a distance from the DLA of $<$ 20 h$^{-1}$ kpc. In
this case the size of the emission-line region corresponds to the area
over which the cloud is optically thick, i.e. is indicative of the size
of a Lyman-limit system. We also consider the relation between
DLAs and Lyman-break galaxies (LBGs). If DLAs are gaseous disks
surrounding LBGs, and if the apparent brightnesses and impact
parameters of the few identified DLAs are representative of the
brighter members of the population, then the luminosity distribution of
DLAs is nearly flat, and we would expect that some $70\%$ of the
galaxy counterparts to DLAs at $z \approx 3$ are fainter than $m_R=28$.

\end{abstract}


\section{Introduction}

The quest for non-active galaxies at high redshift has gone through a
dramatic change in the last few years. The number of galaxies at z$>$2
found using the Lyman-break technique (Steidel et al., 1996) is now
counted in hundreds, providing information on the global density of
star formation at early epochs, and the evolution of galaxy clustering.
A different perspective on the formation of galaxies is obtained by
studying the population of damped Ly$\alpha$ absorbers (DLAs), the
objects responsible for the strongest absorption lines seen in the
spectra of quasars. The advantage of DLAs is that they provide a wealth
of information on the chemical evolution of galaxies in the form of the
measurement of the metallicity and dust content of the gas (e.g. Lu et
al., 1996, Pettini et al., 1997, Kulkarni and Fall, 1997). The DLAs have
column densities larger than 2x10$^{20}$ cm$^{-2}$, comparable to the
column density for a sightline through the disk of a nearby spiral
galaxy. In addition the total gas content in DLAs is similar to the
mass in stars and gas at the present epoch (Wolfe et al., 1995). For
these reasons DLAs are widely believed to be the gas reservoirs from
which todays spiral galaxies formed.

The relation between DLAs and Lyman-break galaxies (LBGs) is unclear at
present because very few DLAs have been unambiguously detected in
emission. At low and
intermediate redshifts $z<1$ a number of candidate counterparts of DLAs
have been discovered in deep images (Steidel et al. 1994, 1995, LeBrun
et al.  1997). These candidates display a mix of morphological types
from spiral galaxies to very compact objects (LeBrun et al.  1997), but
so far none has spectroscopic confirmation. At high redshift there are
three DLAs which have been imaged and for which confirmatory
spectroscopy exists; the DLA at z$_{abs}$=2.81 towards PKS0528--250
(M\o ller \& Warren, 1993b), the DLA at z$_{abs}$ = 3.150 towards
Q2233+131\footnote{When comparing DLA emitters to statistical samples of
DLA absorbers, it is important to remember that this object does not
meet the N(HI) criterion of a DLA absorber. Statistically speaking this
is a Lyman--Limit System}
(Djorgovski et al., 1996), and the DLA at z$_{abs}$ = 4.10 towards
DMS2247-0209 (Djorgovski, 1998). The first of these has been imaged
with HST (M\o ller \& Warren, 1998)
and has a continuum half light
radius of $0.13\pm0.06$ arcsec, similar to that measured for the LBGs
(Giavalisco et al., 1996).

Information on the gas sizes of DLAs comes from the measured impact
parameters. Combined with the line density of absorbers $dn/dz$ the
space density may be computed. M\o ller \& Warren, 1998 produced a
preliminary estimate of the space density of high-redshift DLAs by
using the impact parameters of confirmed and candidate counterparts. The
data suggest that for $q_{0}=0.5$ the space density of DLA clouds at
$z>2$ is more than five times the space density of spiral galaxies
locally. For $q_{0}=0.0$ there is no evidence as yet that DLA clouds
are more common than spiral galaxies locally. A single measurement of
21cm absorption against an extended radio source, the quasar 0458-020,
by Briggs et al. (1989), provides additional information on the size of
DLAs. Their analysis indicates that the DLA seen in the spectrum of
this quasar has a gas size greater than 8h$^{-1}$ kpc.

It would clearly be useful to image more high-redshift DLAs, to
understand better their relation to LBGs, but also to measure the star
formation rates and connect these to the rate of consumption of gas
(Pei and Fall 1995). In this paper we present observations of a DLA in
the field of the quasar pair Q0151+048A (qA) and Q0151+048B (qB). The
two quasars have very similar redshifts z$_{em}$=1.922 and  1.937
respectively (M\o ller, Warren \& Fynbo 1998),
and an angular separation of 3.27$\pm$0.01 arcsec. The spectrum of
qA contains a DLA at z$_{abs}$ = 1.9342, close to the emission
redshift, and this was the reason we originally chose this system for
study because of the possibility of detecting Ly$\alpha$ emission due
to photoionisation by either of the quasars (see M\o ller \& Warren
(1993a,b) for a detailed discussion of the predicted effect).
In M\o ller, Warren \& Fynbo 1998, we presented a spectrum
of qA which shows an emission line in the DLA trough, detected at
4$\sigma$. The present paper describes follow--up narrow--band and
broad--band imaging observations of this system.

In section 2 we describe the observations and the data reduction. In
section 3 we describe the PSF-subtraction of qA and qB and the
photometry of the objects found in the field. In section 4 we discuss
the results obtained. Throughout this paper we adopt q$_0$ = 0.5,
H$_0$ = 100h km s$^{-1}$
Mpc$^{-1}$ and $\Lambda$ = 0 unless otherwise stated.

\section{Observations and Data Reduction}

The data were obtained with the 2.56-m Nordic Optical Telescope (NOT)
during four nights beginning September 15, 1996.  Observations of the
Q0151+048A,B field were done in three bands: the standard U and I
filters, and a special narrow--band filter manufactured by Custom
Scientific. The narrow--band filter (CS 3565/20) is tuned to Ly$\alpha$
at the redshift of the DLA $z=1.9342$. It has a central wavelength of
$\lambda3565{\rm\AA}$ \ and a full width at half maximum (FWHM) of
20\AA . The peak transmission of the filter is T($\lambda$)$_{max}$ =
0.385 and red leak is less than $10^{-6}$ to 1.2 $\mu$.  The CCD used
was a $1024^2$ back-side illuminated thinned Tektronix with a pixel
scale of 0.1757$\pm$0.002 arcsec and read noise of 5.4$e^{-}$. The 
QE in the near UV for this detector rises from 0.3 at 3500\AA \ to 
0.6 at 4000\AA .

Conditions were photometric during most of the run allowing accurate
calibration data to be obtained. The seeing ranged from 0.80 arcsec to
1.35 arcsec FWHM in the narrow--band frames and from 0.60 arcsec to 1.30
arcsec FWHM in the U and I frames. Integration times were in most cases
4000 seconds for the narrow--band frames, 1000 seconds for the U, and
300 seconds for the I, ensuring that the noise in each frame is
dominated by photon noise from the sky and not by read noise. Between
exposures the telescope pointing was jittered in steps of 2 to 5 arcsec
to minimize the effect of bad pixels. The total integration times in
the different filters are listed in Table 1, together with the measured
seeing in the combined frames. For the calibration we observed Landolt
(1992) photometric standards for the broad--band data, and the HST
spectrophotometric standards GD71, BD284211 and BD254655 (Colina and
Bohlin, 1994) for the narrow--band data.
 
\begin{table}
\begin{flushleft}
\caption[]{Observations of Q0151+048A,B, 1996 Sept 15 -- 18}
\begin{tabular}{ccr}
\hline\noalign{\smallskip}
\multicolumn{1}{c}{Filter}& Combined seeing & Exposure \\
           &     & \multicolumn{1}{c}{sec}    \\
\noalign{\smallskip}
\hline\noalign{\smallskip}
CS 3565/20 & 1.1 & 62700 \\
U          & 0.9 & 12000 \\
I          & 0.8 &  7800 \\
\noalign{\smallskip}
\hline
\end{tabular}
\end{flushleft}
\end{table}
 
Bias and dark frames were firstly subtracted from the data frames.
Twilight sky frames were obtained and were used to flatten the U and I
data directly. For the narrow--band frames however we found that the
instrument setup allowed a small, but significant, amount of light to
reach the CCD without passing through the filter.  This additional
source of background light, which could be either light scattered
around the filter wheel or light leaking through the camera enclosure,
was too weak to be detectable in the broad--band images, but was a very
significant contribution in the narrow--band images. Here we shall
refer to this source as scattered light.
 
The scattered light produced a pattern mostly across the upper two
thirds of the CCD. We applied a correction for this as follows.
Firstly the contribution to the twilight sky frames was measured by
placing a blocking filter in the filter wheel. From these frames we
determined a model of the scattered light by median and Gaussian
filtering. This model was then scaled and subtracted from the flat
frames prior to the creation of the normalised master-flat.  The
scattered light in the data frames did not have exactly the same
structure across the CCD as in the flat-field frames. The reason for
this is not known, but it is possibly due to a different position of
the guide probe. We therefore modelled the scattered light in the data
frames by summing all the frames, having removed all objects by
interpolation and filtering. This model was then scaled to, and
subtracted from, the individual data frames. After flat fielding the
residual large scale variations left were less than 5\%.

The frames for each filter were then divided in two groups depending on
whether the seeing was greater or smaller than 1 arcsec FWHM, in order
to optimize the combination of the frames. Both groups were then
registered by integer pixel shifts to a common coordinate system and
the frames within each group were combined using the optimal
combination code described by M\o ller \& Warren (1993b), which
maximizes the signal--to--noise ratio for faint sources. The model for
the scattered light was included in the calculation of photon noise
from the sky. Finally, the two summed frames for each filter were
combined by weighting by the inverse of the sky variance.
 
All magnitudes quoted in this paper are on the AB system. The
narrow--band data were calibrated directly onto the AB system, and
magnitudes are denoted $n(AB)$. For the U band we determined the colour
equation $u=U+0.26(U-B)$ relating the instrumental magnitude $u$ to the
standard Johnson $U$. The fit to the U colour equation used data for
seven different stars but is not well determined. There was substantial
scatter about the fit near $U-B=0$, $\sim 0.05$ mag., and evidence that
the relation is non--linear. Such results are typical for this band
(e.g. Bessell 1990). The instrumental magnitudes $u$ were converted to
AB magnitudes using the equation $u(AB)=u+0.49$, determined by
integrating the spectrum of the star GD71 over the passband. Here we
have retained the lower case $u$ for the AB magnitude indicating that
the effective wavelength of the filter lies significantly away from the
standard value. The colour term for the I filter is consistent with
zero i.e. $i=I$, and we used the equation $I(AB)=I+0.43$ (Fukugita et
al 1995) to put the magnitudes onto the AB system.  Details of the
sky noise in the combined images are provided in Table~\ref{sky}.
 

\section{Results}

\begin{figure}
 \begin{center} \epsfig{file=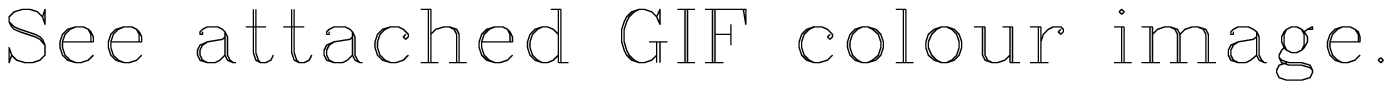,width=8.5cm} \end{center}
 \caption{Field containing the quasar pair Q0151+048A,B. {\it Upper:\,}
 combined I frame, {\it Lower:\,} combined narrow--band frame. The field
 shown is 920x920 pixels, 162x162 arcsec$^{2}$. North is up and east to
 the left. The two quasars are marked qA and qB. The two bright stars
 marked psfA and psfB were used for determination of the point spread
 function. The galaxy marked S5 is discussed in the text (Section 3.3).}
\label{field}
\end{figure}

The field of the quasar pair is illustrated in Fig.~\ref{field} which
shows the final combined I (upper) and narrow--band (lower) images. The
quasars Q0151+048A,B lie near the centre and are marked qA and
qB. Also marked on the image are two bright stars (psfA and psfB) and a
source, named S5, to be discussed later (Section 3.3).

\begin{table}
 \begin{center}
 \caption{Measured rms of sky surface brightness. For the
narrow--band we provide a range due to the presence of the scattered
light gradient across the field.}
 \begin{tabular}{@{}lcccccc}
  passband & rms SB\\
  \hline
           & mag. arcsec$^{-2}$ \\
  \hline
  I(AB)    & 26.9 \\
  u(AB)    & 25.6 \\
  n(AB)    & 25.7-25.8 \\
  \hline
 \end{tabular}
 \label{sky}
 \end{center}
\end{table}


\subsection{Ly$\alpha$ emission near Q$\boldmath 0151+048$}
\begin{figure}
 \epsfig{file=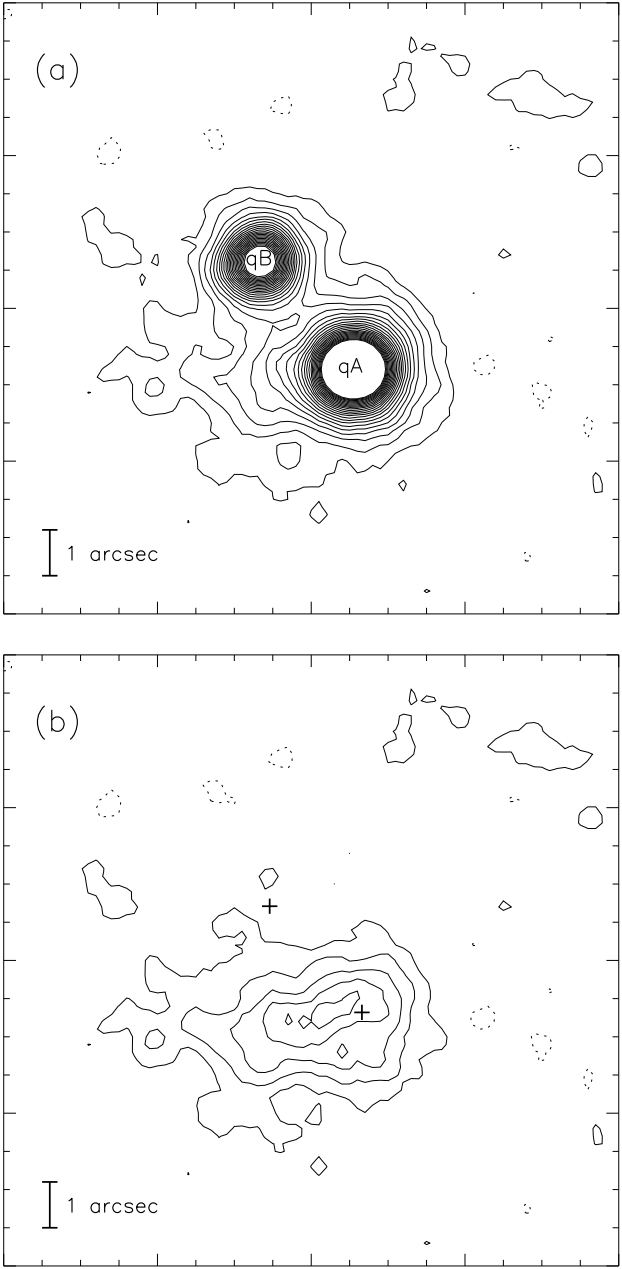,width=8.5cm} \caption{Contour plot of the
 narrow--band frame showing the 80x80 pixels region, 14x14 arcsec$^{2}$,
 surrounding the two quasars. North is up and east to the left. The
 frame has been smoothed by convolution with a Gaussian with $\sigma$
 corresponding to the PSF. {\it Upper:\,} final combined frame showing
 the two quasars marked qA and qB and excess emission east of qA. The
 contour levels are at -3, 3, 6, 9, ... $\times 10^{-17}$ erg
 s$^{-1}$ cm$^{-2}$ arcsec$^{-2}$, with the dotted contours being
 negative. {\it Lower:\,} final combined frame
 (same contour levels as above) after subtraction of the quasar
 images, revealing Ly$\alpha$ emission from the object S4 over a region
 6x3 arcsec$ ^{2}$. The positions where quasar psfs were subtracted
 are marked by crosses.}
\label{fignarrow}
\end{figure}

In the narrow--band image faint emission is seen extending to the east
of qA. To get a clearer view of this in Fig.~\ref{fignarrow}a (upper
panel) we show a contour plot of a small section from the centre of the
narrow--band image. Here, in addition to the images of the two quasars,
a plume of emission, presumably Ly$\alpha$ from the DLA, is clearly
seen extending towards the east of qA. To reveal the full extent of the
Ly$\alpha$ emission we subtracted the images of the two quasars as
follows.
      
The two bright stars marked psfA and psfB in Fig.~\ref{field} were used
with DAOPHOT II (Stetson, 1994) to define the point-spread function
(PSF). The complex image made up of the extended Ly$\alpha$ emitter
super--imposed on the images of the two quasars was decomposed by
iteration. First DAOPHOT II was used to fit, and subtract, the two
point sources. Because of the Ly$\alpha$ emission this will
oversubtract the point sources. We then
used a galaxy model to fit the residuals finding the minimum-$\chi^2$
fit of the model convolved with the psf (as in e.g. Warren et
al. 1996). The model consists of an exponential surface-brightness
profile $\Sigma=\Sigma_s exp(-r/r_s)$, and elliptical shape with
constant ellipticity and orientation. Here $r$ is equal to
$\sqrt{(ab)}$ i.e. the geometric
mean of the semi-major and semi-minor axes. We tried a range of
surface-brightness profiles but the exponential function provided the
best fit. The best-fit galaxy model was then subtracted from the
original image, and a second iteration of DAOPHOT II fitting to the
quasar images followed. After 9 rounds of alternating point source
subtraction and exponential-disc fitting the procedure reached a
stable solution. The frame after subtraction of the images of the two
quasars is shown in Fig.~\ref{fignarrow}b. The parameters of the fit
are $r_s=0.95$ arcsec, a central surface brightness of $3.3\times
10^{-16}$ erg s$^{-1}$ cm$^{-2}$ arcsec$^{-2}$, ellipticity $e=0.53$,
and PA $98^{\circ}$ (E of N). The probability associated with the
$\chi^2$ of the final fit to the Ly$\alpha$ emission is $4\%$. This
rather small value is explained by the asymmetric light profile which
falls off more sharply to the W than to the E.

\begin{figure}
 \epsfig{file=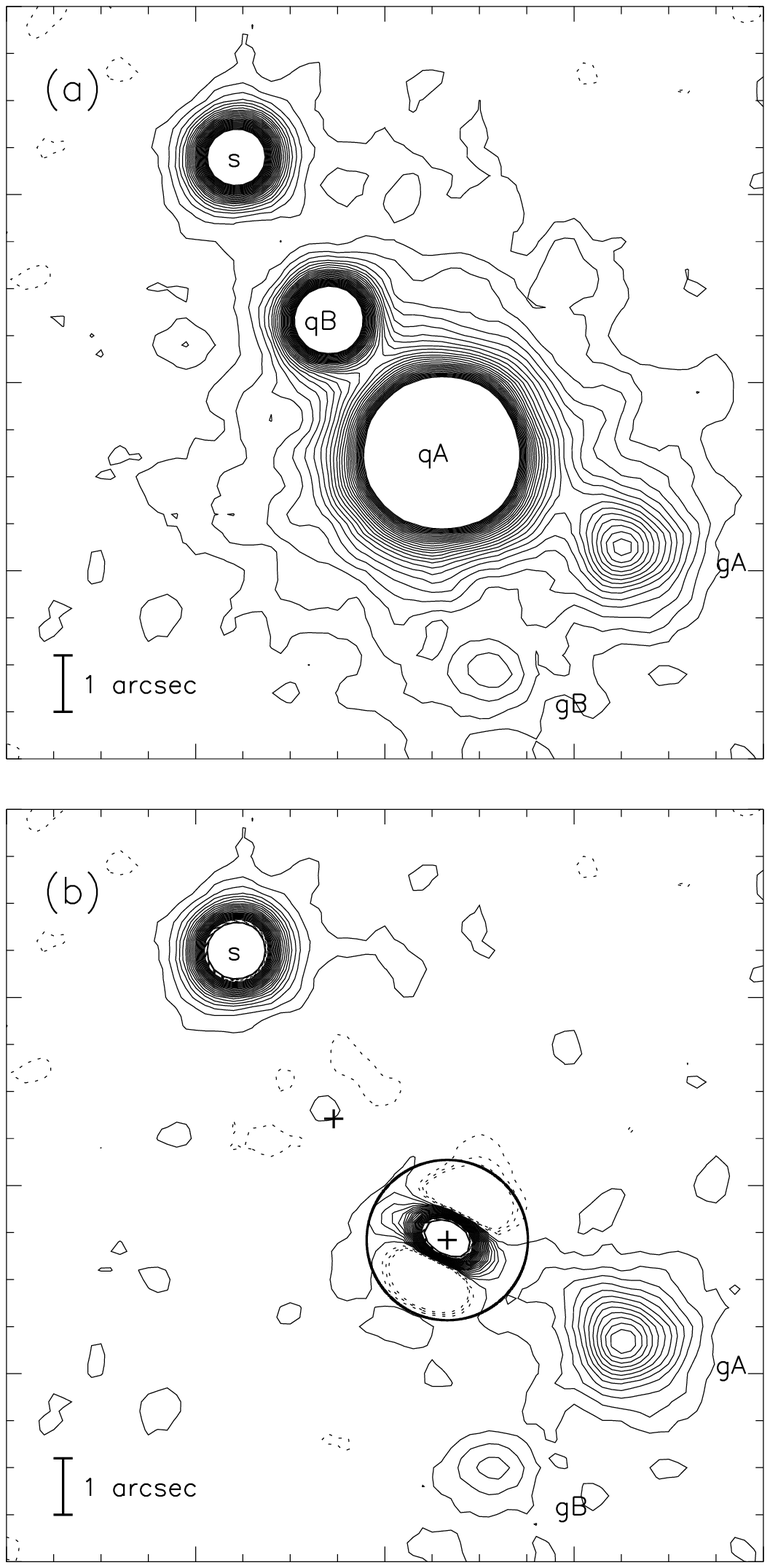,width=8.5cm} 
\caption{I--band contour plot of the same field as in
 Fig.~\ref{fignarrow}. {\it Upper:\,} final combined frame showing, in
 addition to the two quasars, two faint galaxies marked gA and gB, and a
 star marked s.
 The contour levels are -9, -6, -3, 3, 6, ... $\times$
 $1\sigma$ of the skynoise, with the dotted contours being
 negative. {\it Lower:\,} final combined frame
 (same contour levels as above) after subtraction of the quasar images.}
\label{figI}
\end{figure}
\begin{figure}
 \epsfig{file=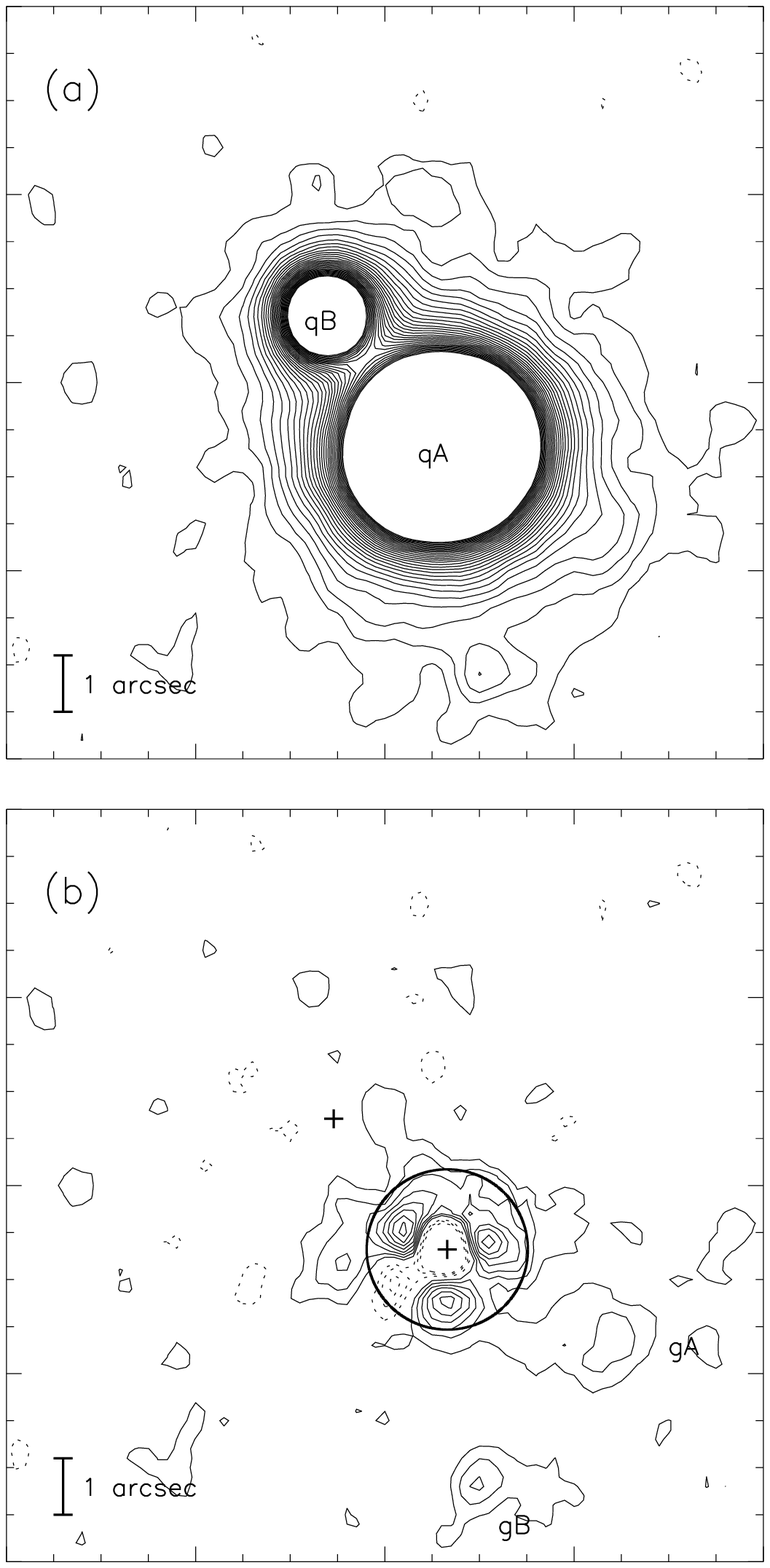,width=8.5cm}
 \caption{U--band contour plot of the same field as in
 Fig.~\ref{fignarrow}. {\it Upper:\,} final combined frame showing
 the two quasars.
 The contour levels are -9, -6, -3, 3, 6, ... $\times$
 $1\sigma$ of the skynoise, with the dotted contours being
 negative. {\it Lower:\,} final combined frame
 (same contour levels as above) after subtraction of the quasar images.}
\label{figU}
\end{figure}

In Fig.~\ref{fignarrow}b emission is seen from an elongated structure
of dimensions approximately 6 x 3 arcsec$^{2}$. In the following we
shall refer to this source of Ly$\alpha$ emission as S4. To perform
photometry of objects in the field we used the package SExtractor
(Bertin \& Arnouts, 1996) and determined {\it isophotal} magnitudes and
so called {\it Automatic Aperture} magnitudes (AAMs), as described
below. The magnitudes of the three objects S4, qA, and qB, were measured
individually in frames from which the images of the other two objects
had been subtracted. The magnitudes of other objects in the vicinity
were measured in frames from which all three images had been removed.

With the exception of S4 the apertures for the {\it isophotal} magnitudes
were set by reference to the frame formed by summing the combined
frames for the three passbands. This frame was convolved with a
Gaussian detection filter with FWHM equal to the seeing. The isophotes
set at 1.5 times the noise measured in the unconvolved frame defined the
apertures and the same apertures were then applied to each passband.
The detection limit for faint objects was set at a minimum aperture size
of 5 pixels (total). Hence, the {\it isophotal} magnitudes provide 
photometry appropriate for accurate colour determination.
For S4 the same procedure was followed, but only
the narrow--band frame was used. Finally, since the catalogue is defined
by detection in the three-passband combination frame, we also ran
SExtractor on the narrow--band frame and found that, with the exception
of S4, no source detected in the narrow--band frame at $S/N>4$ had been
overlooked.

The AAM is intended to give the best estimate of the total flux and is
measured using an elliptical aperture with minor axis $b = 2.5 r_1
\epsilon$ and major axis $a = 2.5 {r_1 \over \epsilon}$, where $r_1$
is the first moment of the light distribution and $\epsilon$ is the
ellipticity. The isophotal narrow--band magnitude and the AAM for S4
correspond to fluxes of
$9.0\pm0.4  \times 10^{-16}$ and
$1.91\pm 0.07 \times 10^{-15}$ erg s$^{-1}$ cm$^{-2}$
respectively. The AAM for S4 (n(AB)=$19.95\pm 0.04$)
agrees well with the total magnitude of the fitted galaxy model
(n(AB)=20.01).

\subsection{Broad--band photometry of S4}

\begin{small}
\begin{table}
\begin{center}
\caption{Photometric properties of the quasars qA and qB, of the faint
galaxies S4, S5, gA and gB, and of the star s. Magnitudes of 
extended objects have been determined in two apertures (isophotal 
and AAM, both defined in Sect. 3.1). For the point sources we
only provide AAM. Note that magnitudes marked $^a$ are not true 
AAM magnitudes, they were determined via scaling of the fitted 
exponential--disk model as detailed in Sect. 3.2. Lower limits 
to magnitudes (upper limits to fluxes) are 2$\sigma$.
  For gA and gB the determination of reliable apertures for AAM was 
impossible due to the nearness of strong residuals of the QSO 
PSF-subtraction.}
\begin{tabular}{l r r rrrr}
Object & u(AB)    & I(AB)    & n(AB)      & Apert.\\
       &          &          &            & (pix)\\
       \hline
       Isoph. &  &  &  & \\
       \hline
       S4 &                &                & $20.76\pm0.05$ & 254\\
       S5 & $24.60\pm0.10$ & $24.04\pm0.09$ & $24.1\pm0.2$   & 37  \\
       gA & $24.28\pm0.12$ & $22.00\pm0.04$ & $>24.0$        & 170 \\
       gB & $26.0\pm0.2$   & $24.88\pm0.14$ & $>25.3$        & 18 \\
       \hline
       AAM &  &  &  & \\
       \hline
       qA & $17.90\pm0.02$ & $17.49\pm0.02$ & $19.72\pm0.04$ & 241 \\
       qB & $21.07\pm0.03$ & $20.87\pm0.02$ & $20.04\pm0.03$ & 241 \\
       s  & $>25.6$        & $20.98\pm0.02$ & $>23.8$        & 341 \\
       S4 & $22.9^a\pm0.2$ & $22.8^a\pm0.3$ & $19.95\pm0.04$ & 690 \\
       S5 & $24.01\pm0.09$ & $23.70\pm0.11$ & $23.6\pm0.2$   & 241 \\
       \hline
       \end{tabular}
       \end{center}
       \end{table}
       \end{small}


Fig.~\ref{figI}a shows a contour plot of the I band image of the
same field as in
Fig.~\ref{fignarrow}. Here we see the two quasars, a star north--east
of the quasars which was not visible in the narrow band image, and two
faint galaxies marked gA and gB south--west of the quasar qA.  In
Fig.~\ref{figI}b psfs have been subtracted at the positions of the two
quasars, leaving behind only the star and the two faint galaxies.
In Fig.~\ref{figI}b we have also overlaid a circle of radius 1.5
arcsec centred on the position of qA. The strong residuals from the
psf--subtraction prevent any detection of additional faint sources
within this region.

In Fig.~\ref{figU}a we show the U--band contour plot of the same region,
and in Fig.~\ref{figU}b again the field after subtraction of the quasar
images. Also in Fig.~\ref{figU}b we have overlaid a circle of radius 1.5
arcsec centred on the position of qA.
In the psf--subtracted U image we see the two faint galaxies gA and gB,
but in addition there is a faint extension about 2 arcsec east of qA,
roughly at the position of S4.

The narrow band image shows that S4 is extended and that it covers the
bright quasar qA. Therefore, when observed through a broad band filter
the tail of the psf of qA extends across most of
S4 and will, even under good seeing conditions, make the determination
of broad--band magnitudes of S4 an extremely difficult task. Simple
aperture photometry, after fitting and subtracting the quasar psf, is
not possible because of the strong residuals from the psf--subtraction,
and because there is a degeneracy in the determination of how much
of the faint extended flux is due to the tail of the quasar psf,
and how much is due
to S4. In order to break this degeneracy, and to obtain an objective
measure for the broad--band magnitudes of S4 we proceeded as follows:

On the assumption that the broad--band flux can be fitted by the
exponential--disk model defined by the fit to the narrow--band flux, we
firstly subtracted a scaled narrow--band model from the broad band
image. We then subtracted a psf at the position of qA. The quasar psf
was scaled so that the integrated residual flux measured in a
rectangular aperture (5.6 arcsec by 3.3 arcsec) covering both the quasar
and the central region of S4, was exactly zero. This calculation was
performed for a range of values of the scaling of the model for S4, and
for each combination of S4--model and quasar psf the total $\chi^2$ of
the fit to the data was calculated. The I and U band magnitudes of S4
provided in Table 3 are those corresponding to the quasar--psf/S4--model
combinations providing the minimum $\chi^2$. The errors
quoted are those where $\chi^2$ increased by one. The rectangular
aperture was defined so as not to include flux from the galaxy gA. The
$\chi^2$ was calculated in a smaller aperture to exclude the central
part of the quasar psf where it would be dominated by noise
from the strong central psf--residuals.
Table 3 provides a summary of the photometry of the object S4, both
quasars, and several other objects found in the field.

M\o ller \& Warren, 1998
found evidence that a DLA--galaxy at z=2.81, as well as its
two companion galaxies, all had one (or more) compact cores of
continuum--emission, while the Ly$\alpha$ emission from the same
objects extended over larger areas. If this is a general feature of
high redshift galaxies we could expect to see one (or several)
continuum emitting cores in S4. Our current data are not adequate to
resolve this question. Clearly higher resolution and deeper
broad--band imaging is required.

\subsection{The impact parameter b$_{\bf DLA}$ of S4}

In M\o ller \& Warren (1998) the DLA impact parameter b$_{\rm DLA}$
was defined as the projected
distance between the line--of--sight to the quasar to that of the
centroid of the continuum flux of the DLA absorber. In the case of
Q0151+048 we clearly see extended Ly$\alpha$ emission from the DLA
absorber, and our analysis of the broad band data shows that they are
not inconsistent with a broad--band flux distribution identical to
that of the Ly$\alpha$ flux. The centroid of the model fit to
the narrow--band data is located 0.93 arcsec east, 0.03 arcsec south
of Q0151+048A (b$_{\rm DLA,Ly\alpha}$ = 0.93 arcsec), and we shall
for now adopt this as the impact parameter of the DLA in Q0151+048A.

\subsection{Ly$\alpha$ in the surrounding field}

\begin{figure}
 \begin{center}
 \epsfig{file=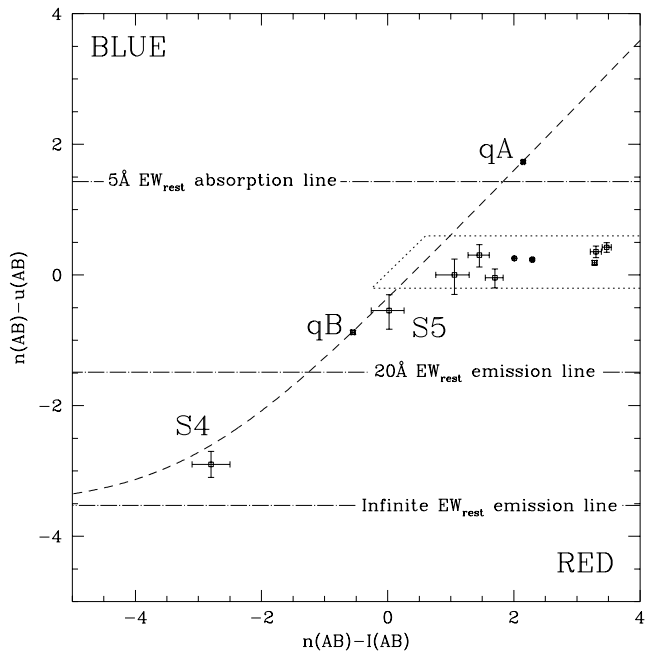,height=9cm}
 \end{center}
 \caption{ Two-colour diagram $n(AB)-u(AB)$ {\it versus} $n(AB)-I(AB)$
 for the 11 detected objects with signal--to--noise ratio in the
 narrow--band frame $>4$. The dashed line is a line of constant
 $u-I$ in the continuum indicating that the two quasars have very 
 similar $u-I$ continuum colours
 but very different Ly$\alpha$ line equivalent widths (EWs).
 The dotted lines confine the expected region of objects with no special
 features in the narrow filter. The horizontal lines are lines of
 constant rest-frame EW=20\AA \  for an emission line at $z=1.93$
 (lower) and rest-frame EW=5\AA \ for an absorption line (upper) 
 that lies in the narrow--band filter.}
\label{NINU}
\end{figure}

An object with Ly$\alpha$ emission at z=1.93 will appear bright in the
narrow--band compared to U and I, so we can use the results of the
photometry to search for other Ly$\alpha$ emitters in the field. In
Fig.~\ref{NINU} we plot the $n(AB)-u(AB)$ {\it versus} $n(AB)-I(AB)$
two--colour diagram for all objects detected at a signal--to--noise
ratio larger than 4 in the narrow--band frame.  Strong Ly$\alpha$
emitters at $z=1.93$ will lie in the lower left region of this diagram.
Conversely an object with absorption in the narrow--band filter will lie
in the upper right corner. 
Since our narrow--band filter is fairly well centred in the U--band
most objects will have zero $n(AB)-u(AB)$ colour, and horizontal lines
in Fig. 5 are to a good approximation lines of constant equivalent
absorption or emission line width.
Lines of constant continuum $U-I$ will appear as
diagonals in Fig.~\ref{NINU} with the bluest objects in the upper left
corner and red objects in the lower right. The dashed curve shows
the location of a series of simulated objects with identical power--law
continuum on which is superimposed absorption or emission lines of
different strengths in the narrow pass--band. The curve mostly falls
along a diagonal of the constant $U-I$ of the continuum, but bends to
the left as the emission line starts to dominate the flux in the
U--band filter. The dashed curve approaches asymptotically a model
with an emission line of infinite equivalent width (dot--dash line).

The quasar qA lies in the upper part of the diagram, as expected
because of absorption by the DLA. Conversely the quasar qB lies in the
lower left part of the diagram. The filter is too narrow to contain all
of the Ly$\alpha$ emission line for the quasar qB, but even so the
$n(AB)-u(AB)$ implies that the mean flux level over the narrow--band
filter is at a level three times greater than the average over the U
band. Therefore the DLA seen in the spectrum of the quasar qA cannot
be present in the spectrum of qB. If the quasar qB also lies behind the
cloud responsible for the DLA, then the column density is much lower at
the position of qB.

There is another candidate Ly$\alpha$ emitter seen in Fig. 5, marked
S5. The location of this candidate is shown in Fig~\ref{field}.
The object is detected at a S/N of 4.8, 14, and 13 in the
narrow, U, and I band images respectively. The corresponding flux of
the candidate Ly$\alpha$ line is
$6.6\pm 1.3 \times 10^{-17}$ erg cm$^{-2}$ s$^{-1}$ and the restframe 
equivalent width of the line is $9$\AA$\pm2$\AA.
The angular separation of S5 from the line of sight to the
quasar is 40.9 arcsec, which corresponds to 169 h$^{-1}$ kpc 
(q$_{0}$=0.5) and 263 h$^{-1}$ kpc (q$_{0}$=0.0) at $z=1.93$.

\section{Discussion}

\subsection{The origin of the large Ly$\alpha$ luminosity}

The Ly$\alpha$ luminosity of S4 is 1.2 x 10$^{43}$ h$^{-2}$ erg
s$^{-1}$. Other high redshift DLAs for which the Ly$\alpha$ flux has
been measured are the DLA at $z=2.81$ towards PKS0528--250 (M\o ller \&
Warren, 1993b), the DLA at z$_{abs}$ = 3.150 towards Q2233+131
(Djorgovski et al., 1996), and the DLA at z$_{abs}$ = 3.083 towards
Q2059--360 (Pettini et al 1995, Robertson and Leibundgut 1997,
Leibundgut and Robertson 1998). The last is a spectroscopic detection
of spatially resolved excess emission in the DLA absorption trough.
The Ly$\alpha$ luminosities for these DLAs are collected in Table 4
and we have assumed
a flux of 2 x 10$^{-16}$ erg s$^{-1}$ cm$^{-2}$ for Q2059--360
based on details in Leibundgut and Robertson (1998).
The measured luminosity for Q0151+048 stands out, being several times
larger than the other values. In this section we consider possible
explanations for this anomaly.

\begin{table*}
 \begin{center} \caption{Measured Ly$\alpha$ luminosities in
 high-redshift DLAs}
 \begin{tabular}{@{}lcccl} quasar & z$_{em}$ & z$_{abs}$ & luminosity & reference \\ \hline
Q0151+048    & 1.92 & 1.93 & 1.2 x 10$^{43}$ h$^{-2}$ & this paper \\
PKS0528--250 & 2.79 & 2.81 & 1.1 x 10$^{42}$ h$^{-2}$ & M\o ller \& Warren
 (1993b) \\
Q2059-360    & 3.10 & 3.08 & 3.7 x 10$^{42}$ h$^{-2}$ & Pettini et al
 (1995), Leibundgut and Robertson (1998) \\
Q2233+131    & 3.30 & 3.15 & 1.2 x 10$^{42}$ h$^{-2}$ & Djorgovski et
 al. (1996) \\
 \hline \end{tabular} \end{center}
\end{table*}

In the light of the dearth of successful detections of emission from
DLAs it is noticeable that of the four high-redshift DLAs listed in
Table 4, three have redshifts close to the quasar emission
redshift. This point was noted by M\o ller et al (1998) who suggested
that photoionisation by the quasar was a plausible explanation for the
Ly$\alpha$ emission for Q0151+048 and Q2059--360, although not for
PKS0528--250 where the continuum emission detected from the DLA is
sufficient to account for the Ly$\alpha$ emission (M\o ller \& Warren,
1998), and where recent spectroscopic observations (Ge et al., 1997)
has ruled out that the quasar could be the source of the ionisation.
In the case of Q0151+048 there are two quasars potentially close to
the DLA galaxy. Q0151+048B is not only close to the line of sight to
the DLA galaxy, it also has a redshift ($1.937\pm 0.005$, 
M\o ller et al 1998) consistent with being identical to that of the
DLA galaxy. For this reason, one may slightly favour qB as the source
of ionization. Q0151+048A, on the other hand, has a redshift which
(if interpreted as due to peculiar motion) indicates that it is moving
towards qB and the DLA galaxy with a velocity of about 1200 km s$^{-1}$.
We now consider the possibility of photoionisation for the Q0151+048 
DLA galaxy in more detail.

The relative distances of qA, qB, and the DLA cannot be determined from
the redshifts because of peculiar velocities. The DLA lies in front of
qA, since it absorbs it, but qB could lie either at larger or smaller
distance than the DLA. If qB is beyond the DLA then the region of 
high-column density of the absorber does not extend across it. This is
because the $n(AB)-u(AB)$ colour of qB corresponds to an average flux
level in the $n$ band of three times the continuum level (Fig 5),
so the Ly$\alpha$
emission line of qB is not strongly absorbed. One possible explanation
for the extended Ly$\alpha$ emission is that qB lies nearer than the
DLA and lights up the near face of the absorber (Fynbo, M\o ller, and
Warren, 1998). However it is also possible that qA is the source of
photoionisation. Since qA under this assumption illuminates the
backside of the absorber, in the simplest picture one would expect to
see a hole in the Ly$\alpha$ emission, containing the line of sight to
the quasar, over the area
where the optical depth is high since over this region the Ly$\alpha$
photons escape out of the back of the absorber. Ly$\alpha$ photons
would only be detectable from an annulus around the DLA where the
optical depth is approximately unity.
However if DLA clouds are sufficiently inhomogeneous on scales much
smaller in extent than the seeing disk, the hole may not be seen as
Ly$\alpha$ photons could travel between the densest regions.

The velocity offset of the emission line relative to the DLA for
Q0151+048 provides an additional clue to the spatial arrangement of the
three objects qA, qB, and S4. The emission line detected for S4
is offset by $\sim300$ km s$^{-1}$ to the blue of the DLA
redshift (M\o ller et al 1998). In the case of Q2059--360 the emission
line lies 490 km s$^{-1}$ to the red of the DLA redshift (Leibundgut
and Robertson 1998). One possible explanation of those
velocity differences is that they are a consequence of
photoionisation by the quasar. The radiation pressure of the resonantly
scattered Ly$\alpha$ photons would cause the HII skin of the DLA to
expand away from the HI in the direction of the quasar (Williams 1972,
Urbaniak and Wolfe 1981). This would provide an explanation of why the
emission line for Q2059--360 is redshifted relative to the DLA. In this
case some of the Ly$\alpha$ photons, produced on the far side, are able
to penetrate the cloud because of the reduced optical depth in the red
wing of the damped absorption line. Those photons would be observed as
a redshifted emission line. For Q0151+048 the blueshift of the emission
line would imply that qB is the source of ionising
photons and must lie on the near side of the DLA.

The above arguments are far from conclusive, but the picture that qB is
the source of ionising photons is at least plausible. We can make this
statement more quantitative by computing how near qB must lie in order
to explain the flux observed. To do this we follow the
analysis presented by Warren and M\o ller (1996, their equations nos 1
and 2). Consider a disk of HI of large optical depth, at distance $d$
from a quasar of absolute magnitude $M_B$, and with normal inclined at
angle $\phi$ to the line from the quasar to the cloud. The surface
brightness of the cloud over the region of high optical depth is given
by:
$$\Sigma_{Ly\alpha}=3.66\times10^{-22}\frac{10^{-0.4M_B}}{d^2}
\frac{cos\phi}{\alpha}(\frac{912}{4400})^\alpha$$
$$ \hbox{} \mathrm{erg\, s^{-1}cm^{-2}arcsec^{-2}}$$ where $\alpha$ is
the quasar continuum slope in the power-law representation
$f_\nu\propto\nu^{-\alpha}$, and $d$ is in kpc. Therefore the distance
of the quasar is given by
$$d=\sqrt{3.66\times10^{-22}\frac{10^{-0.4M_B}}{\Sigma_{Ly\alpha}}
\frac{cos\phi}{\alpha}(\frac{912}{4400})^\alpha\:\:\:{\rm kpc}}$$ We
set $\Sigma_{Ly\alpha}$ equal to the peak surface brightness for the
model fit to S4 (\S3) $\Sigma_{Ly\alpha}=3.3\times 10^{-16}$ erg
s$^{-1}$ cm$^{-2}$ arcsec$^{-2}$ i.e. where the surface brightness
peaks the optical depth of HI visible to the quasar is assumed to be
very high, and the drop off in surface brightness from the peak
corresponds to a decline in column density. The absolute magnitude for
the quasar is computed from the apparent magnitude $I(AB)=20.87$, Table
3. The upper limit to the distance of qB from the cloud is then
obtained by setting $cos\phi=1.0$.  For values of $\alpha=0.7$ and 1.0
we find $d<20$h$^{-1}$ kpc and $<14$h$^{-1}$ kpc, respectively, for
$q_0=0.5$. For $q_0=0.0$ the values are $d<32$h$^{-1}$ kpc and
$<23$h$^{-1}$ kpc. These values are all the more plausible
since they are larger than, but similar to, the projected distance from
qB to the DLA, 12h$^{-1}$ and 19h$^{-1}$, for $q_0=0.5$ and $q_0=0.0$
respectively. This means that the Ly$\alpha$ emission can be explained
by photoionisation by qB without having to invoke an unlikely
geometry. For example for $q_0=0.5$, setting $cos\phi=0.7$,
$\alpha=1.0$, leads to $d=14.5$h$^{-1}$ kpc, which corresponds to an
angle $\theta=30^{\circ}$ between the line joining qB to the DLA and
the plane of the sky. The boundary of detectable Ly$\alpha$ emission
corresponds to the point where the optical depth in HI becomes small.
In other words the quasar is highlighting a Lyman-limit system. The
size of the DLA contained would be smaller.

For completeness we consider the possibility that the photoionisation
model is not correct and that the Ly$\alpha$ emission is due to star
formation. Using the Kennicutt (1983) prescription SFR =
L(H$\alpha$)/1.12 x 10$^{41}$ erg s$^{-1}$ and assuming
L(Ly$\alpha$)/L(H$\alpha$)=10 and negligible dust extinction leads to 

$$
SFR = 11(26)h^{-2} M_{\sun} \ yr^{-1} \: q_0=0.5(0.0)
$$

\noindent
However, with this SFR we would expect S4 to have a U-band magnitude
of the order u(AB)=21 from the continuum 
(assuming SFR = L$_{1500}/(1.6\times10^{40} erg\: s^{-1} $\AA$^{-1}$)), 
which is not seen. Hence, star formation within S4 is very 
unlikely to be the source of the ioninising photons.

\subsection{The relation between DLAs and Lyman-break galaxies}

At high redshift $z\sim3$ the comoving mass density in neutral gas in
the DLAs is similar to the mass density in stars today. In addition the
global star formation rate in the DLAs at high redshift predicted from
the rate of decline of the cosmic density of HI (Pei and Fall 1995) is
in reasonable agreement with the measured global SFR in the Lyman break
galaxies (Madau et al 1996), given the uncertainties in the two
estimates. It is natural therefore to consider the connection between
the two populations, DLAs and LBGs. To address this issue we will make
the assumption that every LBG is surrounded by a DLA disk and ask
whether it is then possible to reproduce the measured properties of the
high-z DLA population.

A summary of the status of the search for LBGs was given by Dickinson
(1998). Fig 9 in Dickinson (1998) provides the current best estimate of
the luminosity function (LF) of galaxies at $z \approx 3$, for
$q_0=0.5$. This estimate is quite uncertain for two reasons. Firstly it
not yet clear how complete the LBG samples are, i.e. how many
high-redshift galaxies fall outside their colour selection
criteria. Secondly the faint-end slope of the LF, i.e. $\alpha$ in the
Schechter parameterisation, is not well determined, relying as it does
on the relatively small number of faint LBGs found in the Hubble Deep
Field. Because of the small field of view of the HDF, and the
possibility of normalisation errors due to clustering, Dickinson (1998)
rescaled the HDF numbers to tie smoothly on to the (well-determined)
numbers at brighter magnitudes. For the present we will assume that the
Schechter fit to the LF present by Dickinson (1998) is an adequate
representation of the LF of LBGs.

Imagine now that every LBG lies at the centre of a DLA disk. To relate
the LBG LF to observables of the DLA population we will suppose that
the size of the DLA disk $R_{gas}$ (i.e. the radius beyond which the
column density falls below 2 x 10$^{20}$ cm$^{-2}$) is related to the
galaxy luminosity by a power-law relation of Holmberg form
$R_{gas}/R_{gas}^*=(L/L^*)^t$. The value of $R_{gas}^*$ is fixed so as
to reproduce the observed line density of DLAs $dn/dz$, i.e. the
average number of DLAs per unit redshift along a line of sight to a
quasar. For randomly inclined disks the relation between $dn/dz$ and
$R_{gas}^*$ is (Wolfe et al. 1986, eqn nos 7, 8):
$$\frac{dn}{dz}=\frac{\pi}{2}\frac{c}{H}\Gamma(1+2t+\alpha)\phi^*(1+z)^{1/2}
R_{gas}^{*2}\:\:\: (q_0=0.5)$$
where $\Gamma$ is the gamma function.
For randomly inclined disks of radius $R$ the average impact parameter
is $\bar{b}=0.524 R$. Therefore
$$\bar{b}^*=0.524\sqrt{\frac{dn}{dz}\frac{2H}{\pi
c\Gamma(1+2t+\alpha)\phi^*(1+z)^{1/2}}}\:\: (q_0=0.5)$$ At $z=3$
$dn/dz=0.27$ (Wolfe et al., 1995). Together with the Holmberg relation
this provides the expected average impact parameter as a function of
apparent magnitude, at $z=3$, and the expected apparent magnitude
distribution of a sample of DLAs selected by gas cross section. Using
Dickinson's LF these results are plotted in Fig. 6, for different
values of the parameter $t$.

\begin{figure}
 \begin{center} \epsfig{file=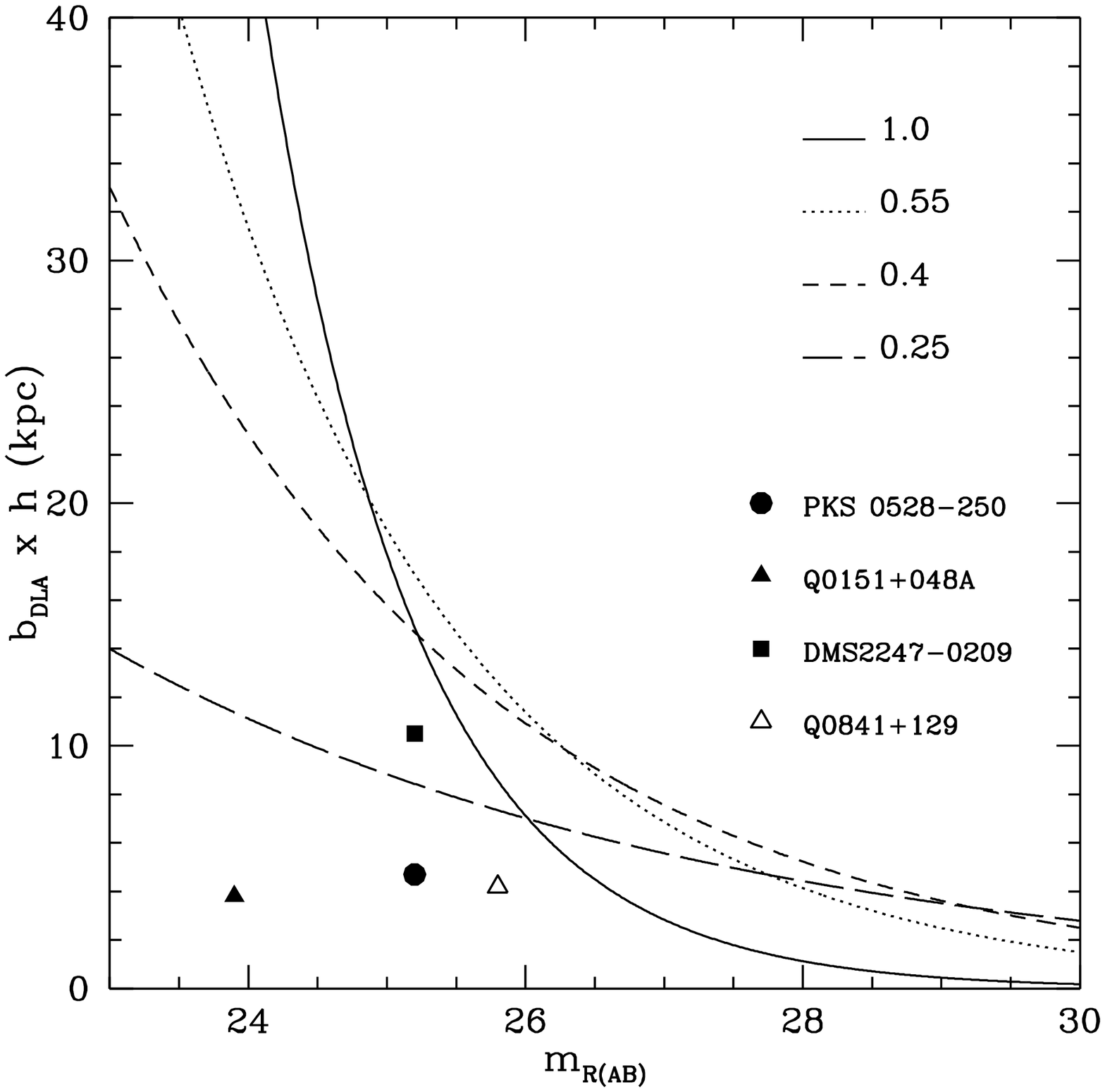, width=9.0cm}
 \epsfig{file=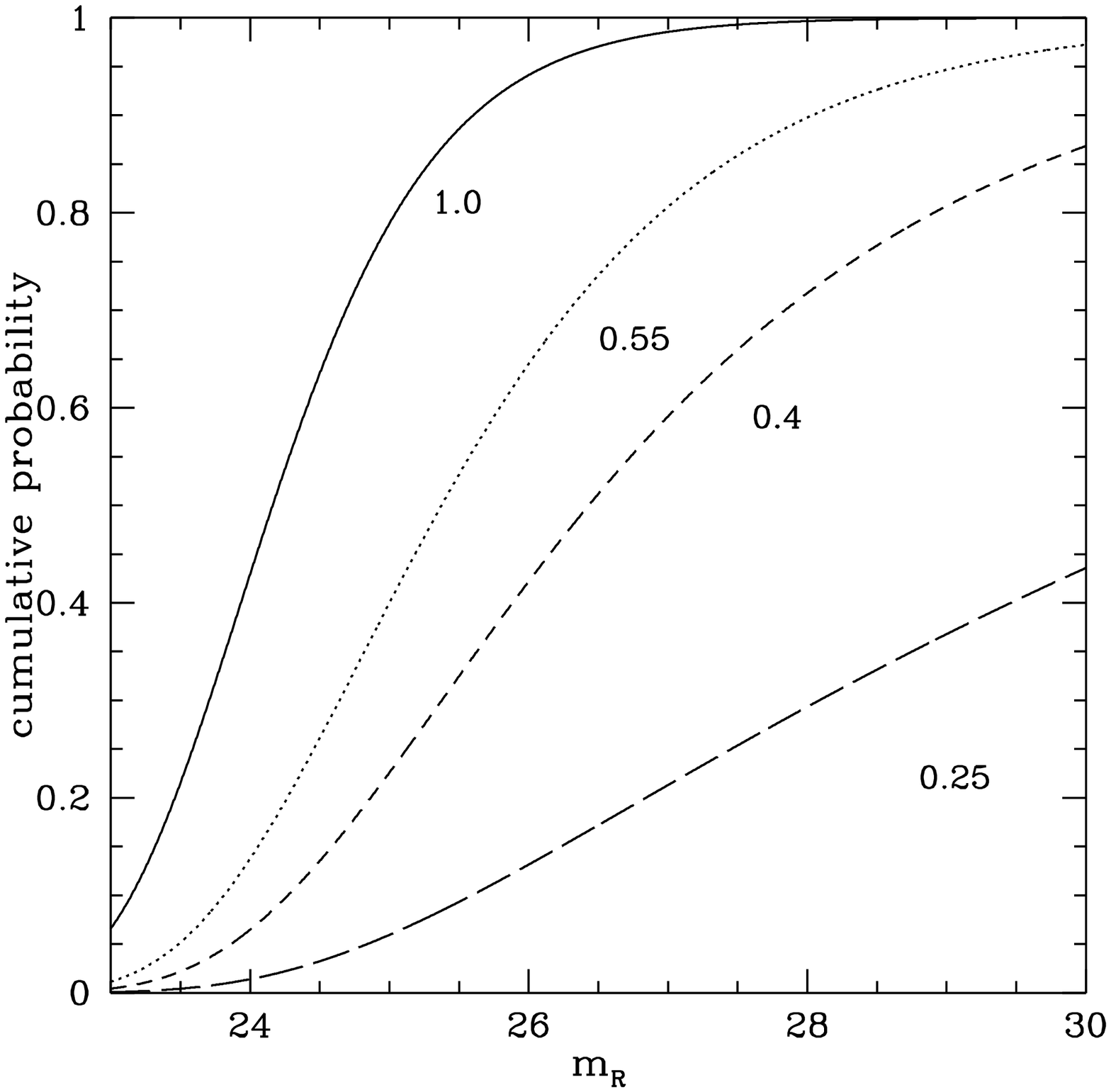, width=9.0cm} 
\caption{
Predicted properties of
the DLA population at $z=3$ if DLAs are gaseous disks surrounding
Lyman-break galaxies. {\it Upper:\,} the curves show predicted values
of the mean impact parameter from the LBG to the line of sight to the
quasar for different values of the Holmberg parameter $t$, as a
function of apparent magnitude. The points are measured impact
parameters for the three known DLA galaxies (and one likely candidate)
for which the continuum magnitude of the
galaxy has been measured. {\it Lower:\,} the curves show the predicted
cumulative probability distribution for the apparent magnitudes of the
LBG counterparts to the DLAs
}
\label{DLALBG}
\end{center}
\end{figure}

When using the predictions presented in Fig. 6 it is important to
remember that both the normalization and faint end extrapolation
of the LBG LF are at present poorly known. Also $dn/dz$ for DLAs could
be wrong by a significant factor, in case systematic effects due to
lensing and/or dust obscuration by the DLAs are biasing the number
counts. Nevertheless it is instructive to compare the predicted curves
in Fig. 6 to the few detections currently available and in Fig. 6 we
therefore also show the apparent magnitudes and impact parameters
of the three confirmed DLA galaxies, and a single DLA galaxy candidate,
from which continuum emission has been detected. Apparent magnitudes
of the four objects have been transformed to $z=3$ to enable comparison
with the $z=3$ LBG luminosity function.

We now compare properties of the known DLA galaxies against the
predictions, to draw preliminary
conclusions about the parameter $t$. It should be noted firstly that
faint DLAs with small impact parameters would be very difficult to
detect, so that the region $m_R>26$ and $\bar{b}<4$h$^{-1}$ is so far
largely unexplored. On the other hand DLAs brighter than $m_R=25$ at
impact parameters $\bar{b}>10$h$^{-1}$ kpc would be detectable from the
ground. If $t=1.0$ most of the LBG counterparts to DLAs would lie in
this region, but the few actual detections are fainter and have small
impact parameters. Therefore large values of $t$ are not compatible
with the observations.

For nearby spirals a value of $t\sim0.4$ is measured (Wolfe et al.,
1986). However, if the measured impact parameters of the few
high-redshift DLAs detected are representative of the fraction of DLAs
brighter than $m_R=26$, a lower value of $t\sim0.25$ is
indicated. Referring now to the lower diagram of Fig. 6, this would
have the
consequence that $\approx 70\%$ of the LBG galaxy counterparts to DLAs
at $z \approx 3$
are fainter than $m_R=28$. This conclusion is in fact relatively
insensitive to the (poorly-determined) faint-end slope of the LBG
LF. For example if the actual value is either flatter or steeper than
Dickinson's measured value, the curves in the upper part of Fig. 6 are
different, and the value of $t$ that passes through the data points is
different. Nevertheless the corresponding curve in the lower plot for
that value of $t$ is quite similar to the lower curve shown.

In the nearby universe a sample of galaxies selected by gas cross
section has a luminosity distribution peaked near $L*$. The reason why
this is probably not the case at high-redshift can be seen as
follows. For a Schechter representation of the luminosity function the
average luminosity of a sample of galaxies selected by gas cross
section is
$\bar{L}=L^*\Gamma(2+\alpha+2t)/\Gamma(1+\alpha+2t)$. Locally the
canonical numbers for spirals are $\alpha=-1.25$ and $t=0.4$ (Lanzetta
et al 1991), yielding $\bar{L}(0)=0.55L^*(0)$. At high redshift taking
$\alpha=-1.38$ and $t=0.25$ gives a much fainter average luminosity
$\bar{L}(z)=0.1L^*(z)$. This is a consequence of the fact that at high
redshift the faint end slope of the luminosity distribution $\alpha+2t$
is closer to the critical value $-1.0$, which corresponds to zero average
luminosity. 

\section*{Acknowledgments}
Nordic Optical Telescope is operated on the island of La Palma
jointly by Denmark, Finland, Iceland, Norway, and Sweden, in the
Spanish Observatorio del Roque de los Muchachos of the
Instituto de Astrofisica de Canarias.

J.U.F. wishes to thank B. Thomsen for many helpful discussions during
preparation for the observing run and is grateful to E. Bertin for
guidance in the use of SExtractor.  J.U.F. and S.J.W. gratefully
acknowledge support from the STScI and ESO visitors programmes,
respectively. We thank our referee, Bruno Leibundgut, for many valuable
comments which helped us clarify several points in the manuscript.

\bsp

\label{lastpage}

\end{document}